\documentclass[%
 reprint,
superscriptaddress,
 amsmath,amssymb,
 aps,
 prb,
 floatfix,
]{revtex4-2}

\usepackage{graphicx}% Include figure files
\usepackage{dcolumn}% Align table columns on decimal point
\usepackage{bm}% bold math
\usepackage{physics}
\usepackage{upgreek}
\usepackage{gensymb}
\usepackage{units}
\usepackage{float}
\usepackage[colorlinks=true,linkcolor=blue,urlcolor=blue,citecolor=blue]{hyperref}

\usepackage[commentmarkup=uwave,deletedmarkup=sout]{changes}
\definechangesauthor[name=Malte Grosche, color=blue]{fmg}

\begin{document}
\title{Metamagnetic ripples in the UTe$_2$ high magnetic field phase diagram}

\author{Zheyu Wu}
\author{Hanyi Chen}
\author{Mengmeng Long}
\affiliation{Cavendish Laboratory, University of Cambridge,\\
 JJ Thomson Avenue, Cambridge, CB3 0US, United Kingdom}

 \author{Gangjian Jin}
 \author{Huakun Zuo}
\affiliation{Wuhan National High Magnetic Field Center, Wuhan 430074, China}

\author{Daniel~Shaffer}
\affiliation{Department of Physics, University of Wisconsin-Madison, Madison, Wisconsin 53706, USA}

\author{Dmitry V. Chichinadze}
\affiliation{National High Magnetic Field Laboratory, Tallahassee, Florida, 32310, USA}
\affiliation{Department of Physics, Washington University in St. Louis, St. Louis, MO, 63130 USA}

 \author{Andrej Cabala}
 \author{Vladim\'{i}r~Sechovsk\'{y}}
 \author{Michal Vali{\v{s}}ka}
 \affiliation{Charles University, Faculty of Mathematics and Physics,\\ Department of
Condensed Matter Physics, Ke Karlovu 5, Prague 2, 121 16, Czech Republic}

\author{Zengwei~Zhu}
\affiliation{Wuhan National High Magnetic Field Center, Wuhan 430074, China}

\author{Gilbert G. Lonzarich}
\author{F. Malte Grosche}
\author{Alexander G. Eaton}
 \email{alex.eaton@phy.cam.ac.uk}
\affiliation{Cavendish Laboratory, University of Cambridge,\\
 JJ Thomson Avenue, Cambridge, CB3 0US, United Kingdom}
 
\date{\today}

\begin{abstract}
\noindent
The heavy fermion metamagnet uranium ditelluride possesses two distinct magnetic field-induced superconducting states. One of these superconductive phases resides at magnetic fields immediately below a first-order metamagnetic transition to a field-polarized paramagnetic state at a field strength $H_m$, while the other exists predominantly above $H_m$. However, little is known about the microscopic properties of this polarized paramagnetic state. Here we report pulsed magnetic field measurements tracking the evolution of $H_m$ for polar and azimuthal inclinations of the magnetic field in the vicinity of the crystallographic $b-a$ plane. We uncover a region of the phase diagram at high fields $>$~50~T with a ripple-like non-monotonic dependence of $H_m$ on the orientation of field. Within this ripple in the metamagnetic transition surface, $H_m$ exhibits an anomalous temperature dependence, implying that here the polarized state has a lower entropy than the low field paramagnetic state. Our results point towards the presence of complex magnetic interactions and possible magnetic sub-phases at high magnetic fields in UTe$_2$, which may have important implications for the manifestation of exotic field-induced superconductivity.

\end{abstract}

\maketitle

% The Pauli exclusion principle necessitates that rotationally symmetric superconductors, with an even-parity orbital component of the superconducting pair wavefunction, must have an antiparallel spin configuration~\cite{pauli1925,bauer2012book}. This is the case for the vast majority of known superconductors. By contrast, in a rare number of materials that lack rotational symmetry, pairs may combine with parallel spins and therefore must possess an odd-parity orbital wavefunction component~\cite{Balian-Werthamer1963,APMackenzieRevModPhys.75.657}. This additional degree of freedom in the spin structure can theoretically lead to rich superconducting phase diagrams comprising multiple proximate superconducting phases characterized by distinct order parameters~\cite{leggett_RevModPhys.47.331}.

\section{Introduction}
\label{sec:intro}
The actinoid dichalcogenide UTe$_2$ is a promising candidate for realizing multi-phase odd-parity superconductivity~\cite{Ran2019Science,Aoki_UTe2review2022}, alongside a plethora of other emergent electronic phases~\cite{lewin2023review,qcl,lewin2025halo,Ranfieldboostednatphys2019,knafo2021comparison,Miyake2021,helm2024,LANL_bulk_UTe2,frank2024orphan,tony2025brief,ran2021expansion,wu2026directobservationspilloverhigh,Braithwaite2019,Miyake2019,aoki2020multiple,Thomas2020,LinNevidomskyyPaglione20,Aoki2021,Aoki_Hard,PDWgu2023,CDWaishwarya2023,Kinjo_SciAdv23,Rosuel23,tony2024enhanced,zou2026multiplenodalsuperconductingphases,wu2026electricallycontrollablesuperconductingmemoryeffect,eaton2026cpb,zhu2026evidenceintertwinedpairdensity,yu2026observationkondohybridizationwave,zhu2026topologicalsurfacestatesrevealed,weinberger2026metamagnetismute2rolesitinerancy,IshidaPRB26}. This compound crystallizes in a body-centered orthorhombic structure (space group 71) with $Immm$ symmetry~\cite{Hutanu}. At ambient pressure, three distinct superconducting phases have been observed for various inclinations of applied magnetic field \textbf{H} up to the remarkably high scale of $\mu_0 H \approx$~70~T~\cite{helm2024,lewin2025halo,tony2025brief} for temperature $T\lessapprox2$~K. Evidence favoring the scenario of odd-parity pairing stems chiefly from high critical fields that exceed the Pauli-Chandrasekhar-Clogston limit for all orientations of \textbf{H}~\cite{tony2024enhanced,chandrasekhar1962note,Clogston_PhysRevLett.9.266}, along with only small changes in the Knight shift on crossing the superconducting critical temperature $T_c$ as measured by nuclear magnetic resonance (NMR)~\cite{matsumura2023NMR-aoki,Kinjo23,tokunaga2023longitudinal}.

For \textbf{H} oriented along the crystallographic $b$-axis -- the hard magnetic direction of UTe$_2$ -- superconductivity persists at low temperatures up to $\mu_0 H \approx$~34~T, whereat it is abruptly truncated by a first-order metamagnetic transition to a polarized paramagnetic state~\cite{Ranfieldboostednatphys2019,Knebel2019}. There is strong evidence indicating that the superconductivity observed at $\mu_0 H \gtrapprox$~15~T constitutes a distinct thermodynamic phase from that at lower $H$~\cite{Rosuel23,VasinaPRL25,ValiskaPRL26}. As the orientation of \textbf{H} is rotated away from $b$ the metamagnetic transition field $H_m$ has been reported to increase monotonically~\cite{lewin2023review}. For a narrow angular range of \textbf{H}, yet another superconductive state, referred to as SC3, has been observed at $H > H_m$~\cite{Ranfieldboostednatphys2019}. Initially discovered for rotations from $b$ towards $c$ -- where it typically starts at $\mu_0 H \approx$~40~T and extends up to around 70~T -- recent measurements have indicated that this remarkable field-induced superconductive state occupies an extended three-dimensional section of magnetic field space, encircling the $b$-axis in a toroidal shape~\cite{lewin2025halo,qcl}. At its zenith it exhibits a higher $T_c$ for $\mu_0H >$~40~T than it does at $\mu_0 H = 0$~T~\cite{tony2025brief}. Magnetoconductance measurements to very high $H$ have indicated that this torus of highly magnetophilic superconductivity is anchored to a quantum critical phase boundary demarcating polarized paramagnetism from the normal paramagnetic state~\cite{qcl}. Deciphering the intimate connection between the metamagnetic phase transition and the field-induced superconducting phases is one of the central challenges in understanding the exotic physical properties of UTe$_2$.

\begin{figure*}[t!]
    \includegraphics[width=1\linewidth]{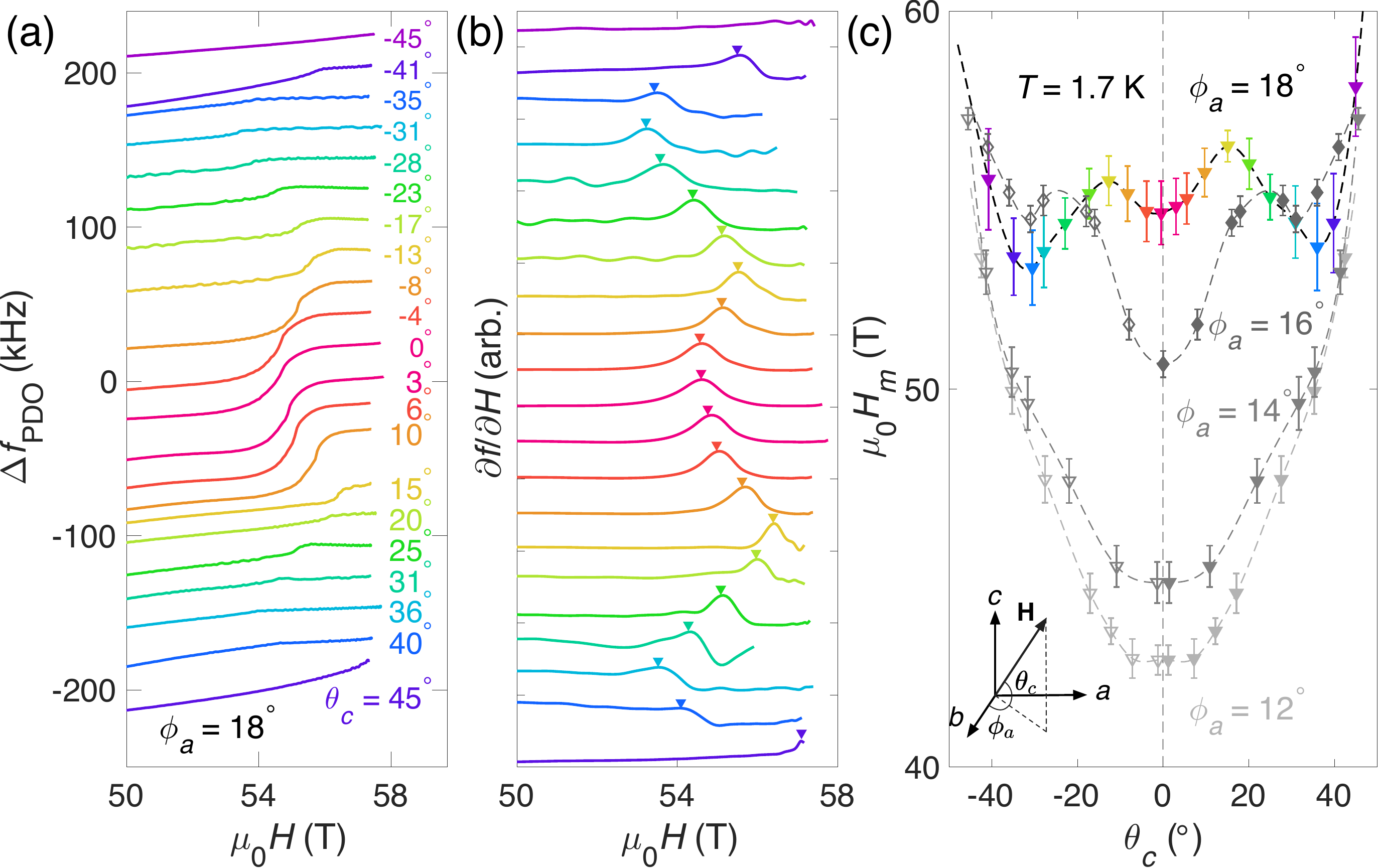}
    \caption{Non-monotonic angular evolution of $H_m$. (a) Contactless conductivity measured by the PDO technique in pulsed magnetic fields of strength $H$ at temperature $T = 1.7$~K. The sample was fixed at an azimuthal inclination of $\phi_a =$~18$\degree$ and rotated through positive and negative values of the polar angle $\theta_c$ up to $\abs{\theta_c} =$~45$\degree$. (b) Derivative with respect to $H$ of the contactless conductivity data in panel (a). Markers identify the location of $H_m$ for each curve, defined as the peak in the derivative of the signal. Curves have been rescaled for ease of comparison. Rotating away from $\theta_c =~$0$\degree$, $H_m$ initially rises before then decreasing to lower values of $H$. (c) $H_m$ versus $\theta_c$ for $\phi_a =$~\{12$\degree$, 14$\degree$, 16$\degree$, 18$\degree$\}. Open points are symmetrized from the solid measured points for $\phi_a =$~\{12$\degree$, 14$\degree$, 16$\degree$\}. Error bars are determined as the standard deviation of a Gaussian fit to $\nicefrac{\partial f}{\partial H}$. The slight asymmetry in $\pm \theta_c$ for $\phi_a = 18\degree$ is likely indicative of a small misalignment of the sample of $\sim 1\degree$. However, within error the profile for the $\phi_a = 18\degree$ data either side of $\theta_c = 0\degree$ broadly agrees very well, exhibiting a clear non-monotonic profile whereby $H_m$ initially rises as $\theta_c$ is tilted away from 0$\degree$, before decreasing to a global minimum around $\abs{\theta_c} = 36\degree$. The inset illustrates the angular coordinate system we use throughout this work, as defined by Eq.~(\ref{eq:Cartesian}).}
    \label{fig:angles}
\end{figure*}

The extent of the high-$H$ superconductivity in the $b-c$ plane has been extensively probed by numerous studies~\cite{lewin2023review,Ranfieldboostednatphys2019,knafo2021comparison,Miyake2021,helm2024,LANL_bulk_UTe2,frank2024orphan,tony2025brief,wu2026directobservationspilloverhigh}. For \textbf{H} tilted out of the $b-c$ plane towards the $a$ direction, the onset field of high-$H$ superconductivity has been reported to track the rise in $H_m$, and its extent has been proposed to extend up to very high values of $H$~\cite{lewin2025halo}. However, few studies have mapped the geometry of the metamagnetic transition surface for large $H_a$ magnetic field components. Given the intimate connection between the metamagnetic transition and the intriguing field-induced superconducting phases of UTe$_2$ -- one of which terminates at $H_m$, while another exists above $H_m$ -- a detailed understanding of the evolution of $H_m$ as a function of magnetic field orientation is necessary to better understand the manifestation of exotic \textbf{H}-induced superconductivity in UTe$_2$.

Prior high-field measurements have indicated that $H_m$ rises monotonically for increasing angular inclination of \textbf{H} away from $b$~\cite{lewin2023review}. Throughout the present study, we will adopt a convention of spherical polar coordinates in which the azimuthal angle $\phi_a$ is defined to be the inclination from $b$ towards $a$, while the polar angle $\theta_c$ is the inclination out of the $b-a$ plane towards the $c$-axis. The  magnetic field components oriented along each of the three Cartesian crystallographic axes are therefore related to these polar and azimuthal inclinations by the following expressions:
\begin{equation}
\begin{aligned}
& H_a=H \cos \theta_c \sin \phi_a \\
& H_b=H \cos \theta_c \cos \phi_a \\
& H_c=H \sin \theta_c.
\end{aligned}
\label{eq:Cartesian}
\end{equation}

Previous studies rotating \textbf{H} in the $b-c$ plane have reported that the evolution of $H_m$ appears to follow a $\nicefrac{1}{\cos{\theta_c}}$ dependence~\cite{helm2024,lewin2023review,Ranfieldboostednatphys2019}. Similarly, for nonzero values of both $\phi_a$ and $\theta_c$, measurements of the metamagnetic transition have discerned a monotonically rising profile of $H_m$ upon increasing either $\theta_c$ or $\phi_a$~\cite{lewin2025halo}.

%By contrast, for nonzero $\phi_a$ and $\theta_c$, measurements of the metamagnetic transition have been found to be well fitted by the phenomenological expression $H_m = H_m^{bc} + \alpha \sin^2(\phi_a) + \beta \sin^4(\phi_a)$, where $H_m^{bc}$ is the value of $H_m$ in the $b-c$ plane (i.e. at the same value of $\theta_c$ for $\phi_a = 0\degree$) while $\alpha$ and $\beta$ are fit parameters~\cite{lewin2025halo}. This gives a monotonically rising profile of $H_m$ upon increasing either $\theta_c$ or $\phi_a$.

\section{Methods}
\label{sec:methods}
Contactless magnetoconductance measurements in pulsed magnetic fields were acquired by the proximity detector oscillator (PDO) technique~\cite{PDO_Altarawneh}. The contactless magnetoconductance of UTe$_2$ has been measured in numerous studies~\cite{Ranfieldboostednatphys2019,LinNevidomskyyPaglione20,ran2021expansion,Eaton2024,tony2024enhanced,frank2024orphan,theo2024,tony2025brief,qcl,weinberger2025pressure,lewin2025halo,weinberger2026metamagnetismute2rolesitinerancy} and is well understood. A sample was mounted on a hand-wound planar coil with 6 turns possessing a diameter approximately matching the length of the sample (to maximize filling factor), with one outermost counter-winding added at a radius calculated to enclose the same amount of magnetic flux as the inner coil (but with opposite polarity). Note that throughout we invert $\Delta f$ so that increasing $f$ corresponds to increasing sample resistance. The PDO circuit rang at a frequency of 32~MHz, which was then mixed down twice to $\sim$1~MHz and acquired using a JYTEK PXIe-69834 data acquisition card sampling at a rate of 30 megasamples per second. The data we present in this study were taken on the downsweep of the magnetic field pulse, which has a slower rate of change of $H$ with time (upsweep data are included in the Appendix). All error bars on $H_m$ are calculated from the standard deviation of Gaussian fitting to the first derivative of the PDO signal with respect to $H$.

Orienting \textbf{H} at nonzero values of $\phi_a$ and $\theta_c$ was achieved by first mounting the PDO coil baseplate atop a small wedge of G10 machined to a desired azimuthal inclination. By rotating the relative orientation of \textbf{H} through the plane orthogonal to the inclination of the G10 wedge, the polar angle $\theta_c$ could therefore be varied at a constant azimuthal inclination of $\phi_a$. The sample was oriented on the PDO baseplate by Laue diffractometry, similar to Refs.~\cite{tony2024enhanced,theo2024}. All data were taken on the same sample, grown by the molten salt flux technique~\cite{PhysRevMaterials.6.073401MSF_UTe2} following the methodology outlined in Ref.~\cite{Eaton2024}, with a resistive $T_c$ (in ambient $H$) of 2.09~K and a residual resistivity ratio of $\sim$~400.

\section{Results}
\label{sec:results}
\subsection{\texorpdfstring{Angular dependence of $H_m$}{Angular dependence of Hm}}
\label{sec:angles}
\begin{figure}[t]
    \includegraphics[width=.9\linewidth]{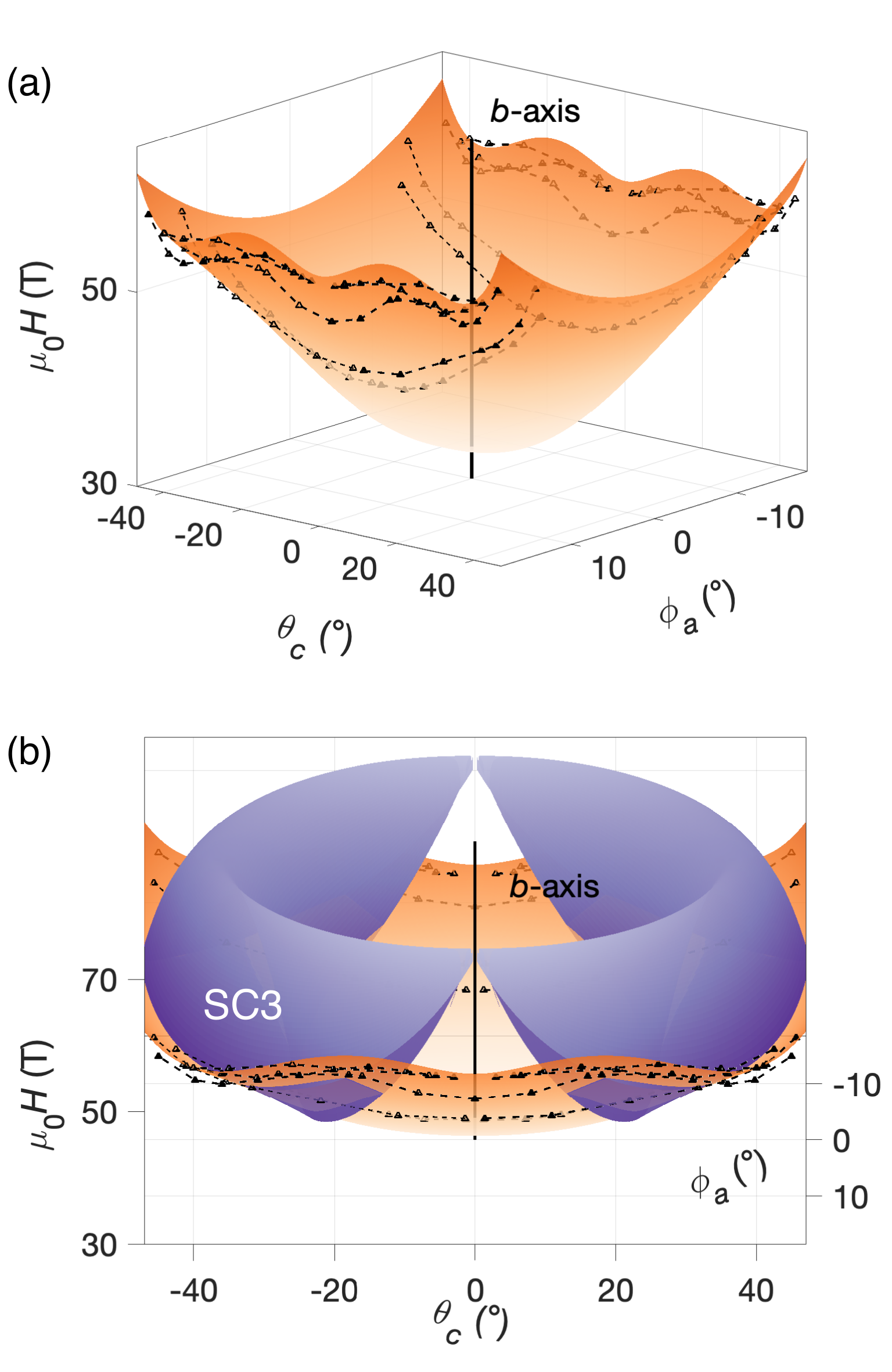}
    \caption{ (a) Ripples in the metamagnetic transition surface of UTe$_2$. Points are reproduced from Fig.~\ref{fig:angles}(c) and plotted here in the three dimensions of $H$, $\theta_c$ and $\phi_a$. The orange surface illustrates the profile of the first-order metamagnetic phase boundary, as measured at a temperature of 1.7~K. For sufficiently high $H$ and $\phi_a$ the evolution of $H_m$ with $\theta_c$ is highly non-monotonic. (b) Rotated view of the rendering from panel (a), with the approximate extent of the toroidal domain of SC3 superconductivity plotted on top (in purple), taken from Ref.~\cite{qcl}. The determined region of SC3 matches closely with that reported in Ref.~\cite{lewin2025halo}. The superconductivity appears to weaken where the metamagnetic ripples are most pronounced.}
    \label{fig:3D}
\end{figure}

\begin{figure*}[t]
    \centering
    \includegraphics[width=0.75\linewidth]{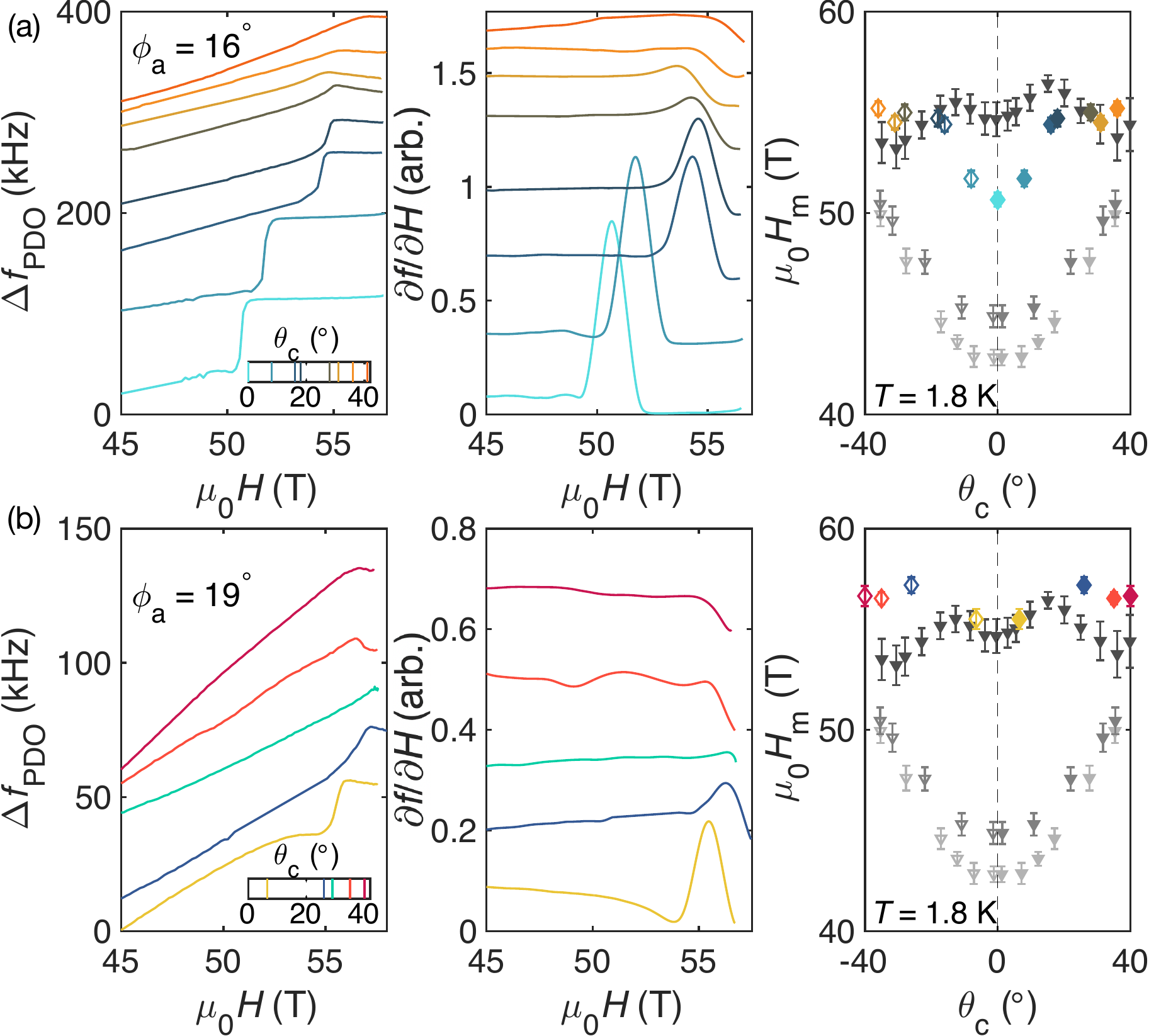}
    \caption{Evolution of the metamagnetic phase landscape under rotations of $\theta_c$ for (a) $\phi_a = 16\degree$ and (b) $\phi_a = 19\degree$. Left: PDO frequency shift $\Delta f_{\text{PDO}}$ versus $H$ for the values of $\theta_c$ indicated by the color scale (inset). Center: the corresponding derivatives $\partial f/\partial H$, offset vertically for clarity. Right: $H_m$ versus $\theta_c$ at $T = 1.8$~K, determined from the peaks in $\partial f/\partial H$ (diamonds, colored as the corresponding curves); open symbols are symmetrized from the measured (solid) points. Gray triangles reproduce the data for $\phi_a = 12\degree$, $14\degree$ and $18\degree$ from Fig.~\ref{fig:angles}(c), with darker shades for larger $\phi_a$. These data were used in part to construct Fig.~\ref{fig:angles}(c).}
    \label{fig:new_rotation}
\end{figure*}

\begin{figure*}[t]
    \centering
    \includegraphics[width=0.75\linewidth]{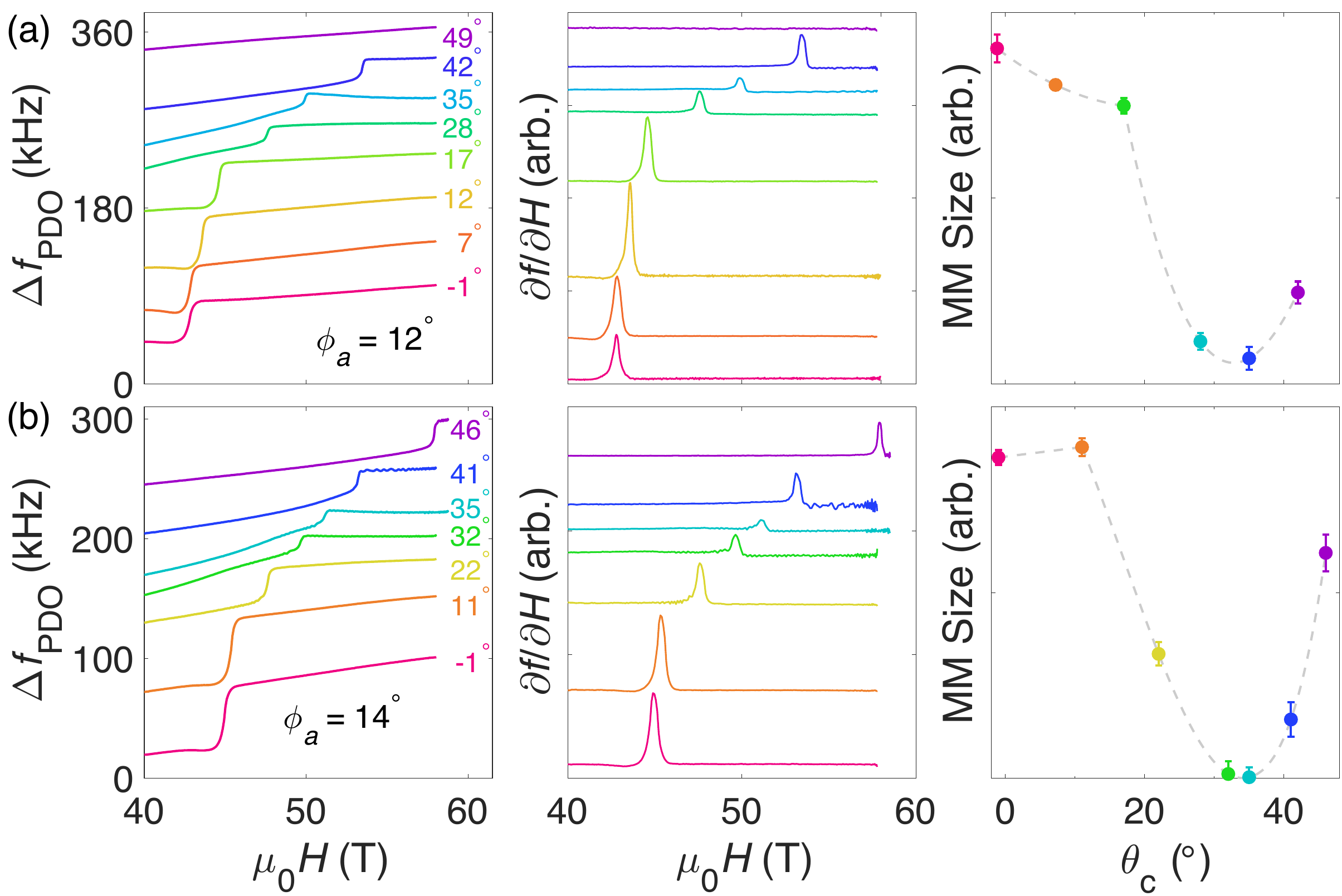}
    \caption{Evolution of the metamagnetic phase landscape under rotations of $\theta_c$ for (a) $\phi_a = 12\degree$ and (b) $\phi_a = 14\degree$. We plot the size of the metamagnetic transition, as indicated by the height of $\partial f/\partial H$. Error bars are determined by 95\% confidence bounds of Gaussian fits to the derivative peaks. The size of the jump follows a non-monotonic dependence on $\theta_c$, diminishing at intermediate orientations when closest to the quantum critical line that bounds the metamagnetic transition surface~\cite{qcl}. The size then increases again at high $\theta_c$, when moving further away again from the quantum critical line.}
    \label{fig:jumps}
\end{figure*}

Here we report measurements of the evolution of $H_m$ in pulsed magnetic fields up to 57~T for rotations through $\theta_c$ at fixed values of $\phi_a$. Figure~\ref{fig:angles} shows the contactless conductivity of UTe$_2$ measured by the PDO technique at incremental $\theta_c$ for $\phi_a =$~18$\degree$. The metamagnetic transition manifests in the PDO signal as a sharp step in the PDO frequency as $H_m$ is crossed. Figure~\ref{fig:angles}(b) identifies the location of $H_m$ for each curve as the maximum of the derivative of the signal. The location of $H_m$ for $\phi_a =$~18$\degree$ and $\phi_a =$~16$\degree$ clearly exhibits a non-monotonic dependence on $\theta_c$. This is in contrast to the behavior observed for $\phi_a =$~14$\degree$ and $\phi_a =$~12$\degree$ (Fig.~\ref{fig:angles}(c)), both of which continuously increase upon tilting $\theta_c$ towards 90$\degree$; that being said, there does appear to be a slight inflection around $\theta_c \approx$~25$\degree$ in the $\phi_a =$~14$\degree$ dataset. Notably, for $\phi_a =$~18$\degree$ we find that $H_m(\theta_c = 36\degree) < H_m(\theta_c = 0\degree)$, with $\mu_0 H_m(\theta_c = 36\degree) = 53(1)$~T and $\mu_0 H_m(\theta_c = 0\degree) = 54.6(8)$~T; for $\phi_a =$~16$\degree$ there is a small local minimum where $\mu_0 H_m(\theta_c = 31\degree) = 54.5(4)$~T, while $\mu_0 H_m(\theta_c = 0\degree) = 50.6(4)$~T.

In Fig.~\ref{fig:3D} we replot the data from Fig.~\ref{fig:angles}(c) -- namely, the locations of $H_m$ determined from rotations of \textbf{H} through $\theta_c$ at fixed $\phi_a =$~\{12$\degree$, 14$\degree$, 16$\degree$, 18$\degree$\} -- this time in the three-dimensional space of $H$, $\theta_c$ and $\phi_a$. The $b$-axis is identified as the solid black line at $\theta_c = \phi_a =$~0$\degree$. The metamagnetic transition surface is drawn in orange, which for sufficiently large $H$ and $\phi_a$ exhibits a distinct ripple-like profile. Interestingly, the SC3 superconductive state appears to diminish around the region where the ripples are most pronounced.

Figure~\ref{fig:new_rotation} shows the PDO data from which the $\phi_a = 16\degree$ profile in Fig.~\ref{fig:angles}(c) was extracted, together with data at $\phi_a = 19\degree$. At $\phi_a = 16\degree$, $H_m$ rises steeply from $\mu_0 H_m \approx 50.6$~T at $\theta_c = 0\degree$ to $\approx 54.5$~T at $\abs{\theta_c} \approx 17\degree$, beyond which the profile flattens, with the shallow local minimum at $\theta_c \approx 31\degree$ noted above. At $\phi_a = 19\degree$ the transition lies between $\mu_0 H_m \approx 55$ and 57~T for all $\abs{\theta_c} \leq 40\degree$, close to the maximum field of our measurements. At this inclination the step in $\Delta f_{\text{PDO}}$ is considerably weaker than at lower $\phi_a$, particularly at large $\abs{\theta_c}$. Measurements to higher fields will be needed to resolve the evolution of the ripple beyond $\phi_a \approx 19\degree$.

The corresponding data for $\phi_a = 12\degree$ and $14\degree$, from which the profiles in Fig.~\ref{fig:angles}(c) were extracted, are shown in Fig.~\ref{fig:jumps}. At both inclinations the metamagnetic transition appears as a sharp step in $\Delta f_{\text{PDO}}$ that moves monotonically to higher $H$ as $\theta_c$ is increased. The height of the corresponding peak in $\partial f/\partial H$, which provides a measure of the size of the metamagnetic transition, is however non-monotonic in $\theta_c$: it diminishes at intermediate inclinations, $\theta_c \approx 30$--$35\degree$, where the metamagnetic transition surface approaches the quantum critical line that bounds it~\cite{qcl}, and grows again at larger $\theta_c$ as the transition moves away from this line.

\subsection{\texorpdfstring{Temperature dependence of $H_m$}{Temperature dependence of Hm}}
\label{sec:tdep}
\begin{figure}[t]
    \includegraphics[width=1\linewidth]{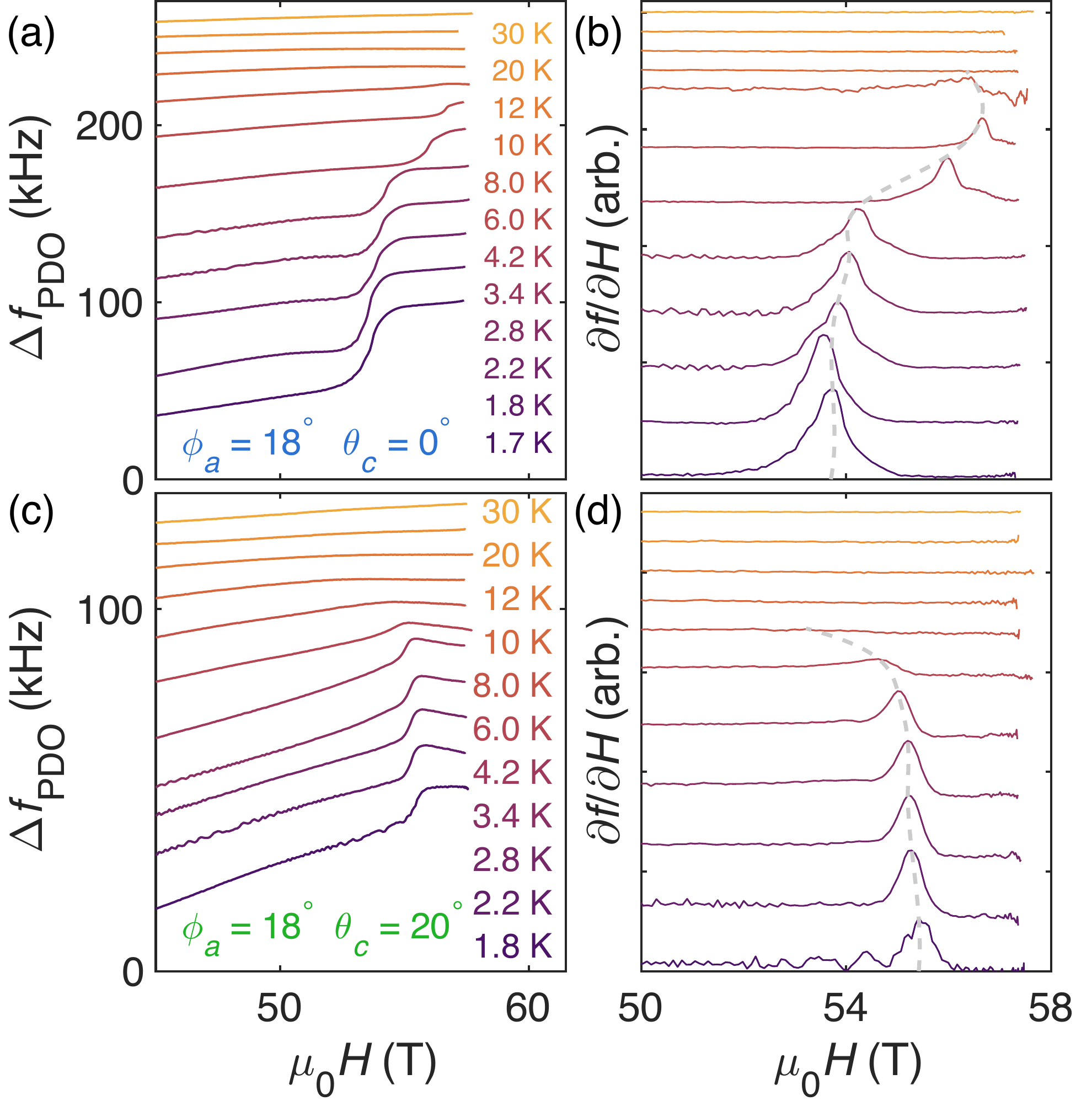}
    \caption{Tracking the temperature dependence of $H_m$. (a) Contactless conductivity data measured by the PDO technique at incremental temperatures for \textbf{H} aligned at $\phi_a = 18\degree$, $\theta_c = 0\degree$, with (b) the corresponding derivatives. (c) PDO measurements for \textbf{H} tilted at $\phi_a = 18\degree$, $\theta_c = 20\degree$, with (d) the corresponding derivatives (again, derivative curves have been rescaled for ease of comparison). Whereas the location of $H_m$ for the $\theta_c = 20\degree$ dataset monotonically decreases with elevated temperature, by comparison at $\theta_c = 0\degree$ $H_m$ initially rises upon heating, before subsequently decreasing.}
    \label{fig:temps}
\end{figure}

\begin{figure}[t]
    \includegraphics[width=1\linewidth]{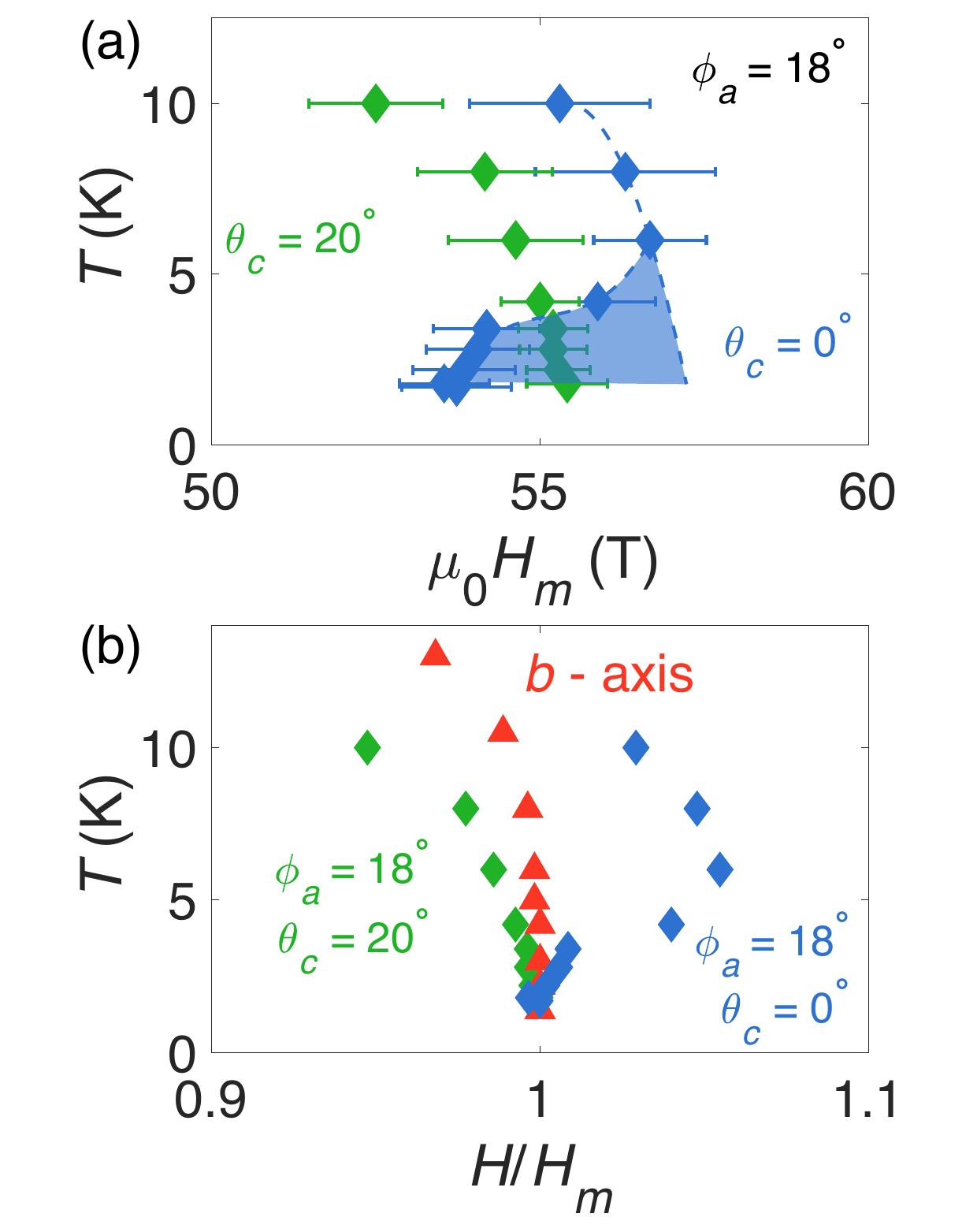}
    \caption{Anomalous temperature evolution of metamagnetism. (a) The values of $H_m$ extracted from the $\nicefrac{\partial f}{\partial H}$ curves in Fig.~\ref{fig:temps} for $\theta_c = 20\degree$ (green) and $\theta_c = 0\degree$ (blue). (b) Normalized $H/H_m$ for the points in panel (a) plotted alongside data for \textbf{H}~$\parallel b$ taken from Ref.~\cite{knafo2019magnetic}. Whereas the $\phi_a = 18\degree$, $\theta_c = 20\degree$ (green) and $\phi_a = \theta_c = 0\degree$ (orange) points monotonically decrease as the temperature is raised, the $\phi_a = 18\degree$, $\theta_c = 0\degree$ (blue) points anomalously increase, reaching a maximum around 6.0~K, before subsequently decreasing at higher $T$.}
    \label{fig:3dtemps}
\end{figure}

\begin{figure*}[t]
    \centering
    \includegraphics[width=0.82\linewidth]{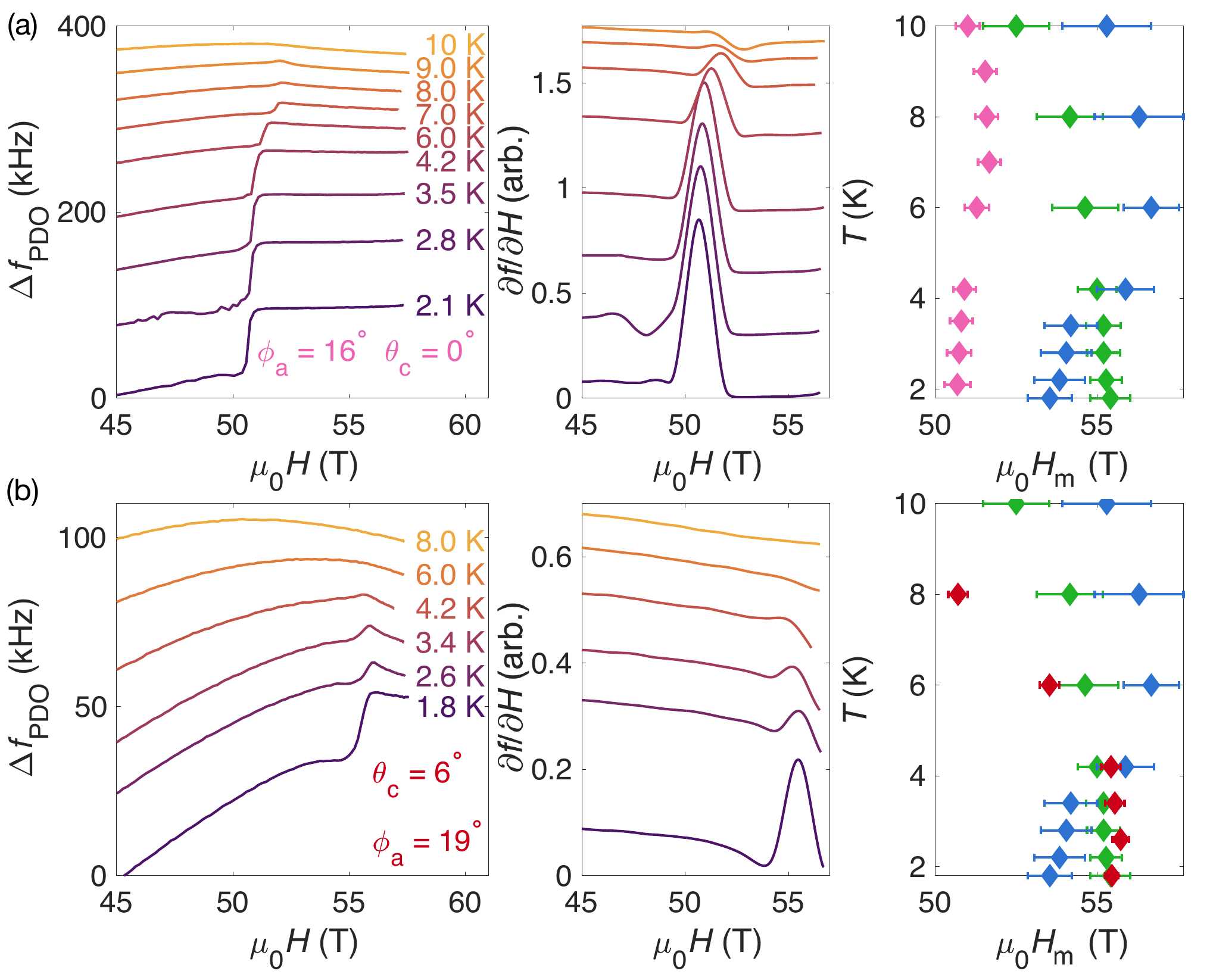}
    \caption{Temperature evolution of the PDO signal for field oriented at (a) $\phi_a = 16\degree$, $\theta_c = 0\degree$ and (b) $\phi_a = 19\degree$, $\theta_c = 6\degree$. Left: PDO signal at incremental temperatures; center: the corresponding derivatives $\partial f/\partial H$; right: $T$ versus $H_m$ for $\phi_a = 16\degree$, $\theta_c = 0\degree$ (pink) and $\phi_a = 19\degree$, $\theta_c = 6\degree$ (red). Both datasets show slight non-monotonicity in $H_m(T)$. The $H_m(T)$ data for $\phi_a = 18\degree$, $\theta_c = 20\degree$ (green) and $\phi_a = 18\degree$, $\theta_c = 0\degree$ (blue) are included for reference.}
    \label{fig:Tdep_new}
\end{figure*}

We track the evolution of $H_m(T)$ for two different angular orientations in Fig.~\ref{fig:temps}. For \textbf{H} oriented at the inclination of $\phi_a = 18\degree$, $\theta_c = 20\degree$ the location of $H_m$ -- as identified by the sharp peak in $\nicefrac{\partial f}{\partial H}$ at low $T$ -- gradually decreases with increasing $T$, and beyond the critical end point of the first-order transition it evolves into a broad crossover. This behavior is consistent with numerous other reports that also studied the evolution in temperature of polarized paramagnetism at high-$H$ in UTe$_2$~\cite{lewin2023review,Miyake2019,knafo2019magnetic,Miyake2021,valiska2024dramatic,qcl}.

By contrast, for \textbf{H} aligned such that $\phi_a = 18\degree$, $\theta_c = 0\degree$ -- at the local minimum of the ripple in $H_m(\theta_c)$ in Fig.~\ref{fig:angles}(c) -- a markedly different behavior is exhibited. At $T =$~1.7~K, $\mu_0H_m =$ 53.7(8)~T. $H_m$ then \textit{increases} upon raising $T$, to a maximal value of $\mu_0H_m =$ 56.7(9)~T at 6.0~K. At higher $T$, $H_m$ then decreases, and the profile of the metamagnetic transition evolves into the broad crossover also observed at other inclinations of \textbf{H}. 

%This anomalous profile of $H_m(T)$ is especially prominent when plotted alongside the normalized locations of $H_m$ for other orientations of \textbf{H}, as we display in Fig.~\ref{fig:3dtemps}b.

The anomalous profile of $H_m(T)$ for $\phi_a = 18\degree$, $\theta_c = 0\degree$ is particularly prominent when plotted alongside the normalized locations of $H_m$ for other orientations of \textbf{H} (Fig.~\ref{fig:3dtemps}(b)). While $H_m(\phi_a = 18\degree$, $\theta_c = 20\degree)$ and $H_m(\phi_a = \theta_c = 0\degree)$ decrease smoothly and monotonically on raising $T$, $H_m(\phi_a = 18\degree$, $\theta_c = 0\degree)$ behaves quite differently. Firstly $H_m$ rises gradually as $T$ is incremented from 1.7~K to 3.4~K, and then rapidly increases around 4~K, reaching its maximum at 6.0~K. At higher $T$ it appears to follow the same trend of other angular orientations, whereby the metamagnetic transition broadens and migrates down towards zero field.

To establish how this anomalous temperature dependence evolves across the ripple, we measured $H_m(T)$ at two further orientations, $\phi_a = 16\degree$, $\theta_c = 0\degree$ and $\phi_a = 19\degree$, $\theta_c = 6\degree$ (Fig.~\ref{fig:Tdep_new}). At $\phi_a = 16\degree$, $\theta_c = 0\degree$ -- which, like $\phi_a = 18\degree$, $\theta_c = 0\degree$, lies at a minimum of $H_m(\theta_c)$ (Fig.~\ref{fig:angles}(c)) -- $H_m$ again increases on warming, by $\approx 1$~T between 2.1~K and $\approx 7$~K, before decreasing at higher $T$. This rise is considerably smaller than the $\approx 3$~T increase observed at $\phi_a = 18\degree$, $\theta_c = 0\degree$. At $\phi_a = 19\degree$, $\theta_c = 6\degree$, $H_m$ rises only marginally between 1.8~K and 2.6~K and then decreases on further warming, more closely resembling the behavior at $\phi_a = 18\degree$, $\theta_c = 20\degree$. Among the orientations studied, the anomalous increase of $H_m$ with $T$ is therefore most pronounced at $\phi_a = 18\degree$, $\theta_c = 0\degree$. For $\phi_a = 18\degree$, the rising- and falling-field sweeps are compared in the Appendix; both show the increase of $H_m$ on warming at $\theta_c = 0\degree$.

\subsection{Landau free energy description}
\label{sec:landau}
\begin{figure}[t!]
    \includegraphics[width=.9\linewidth]{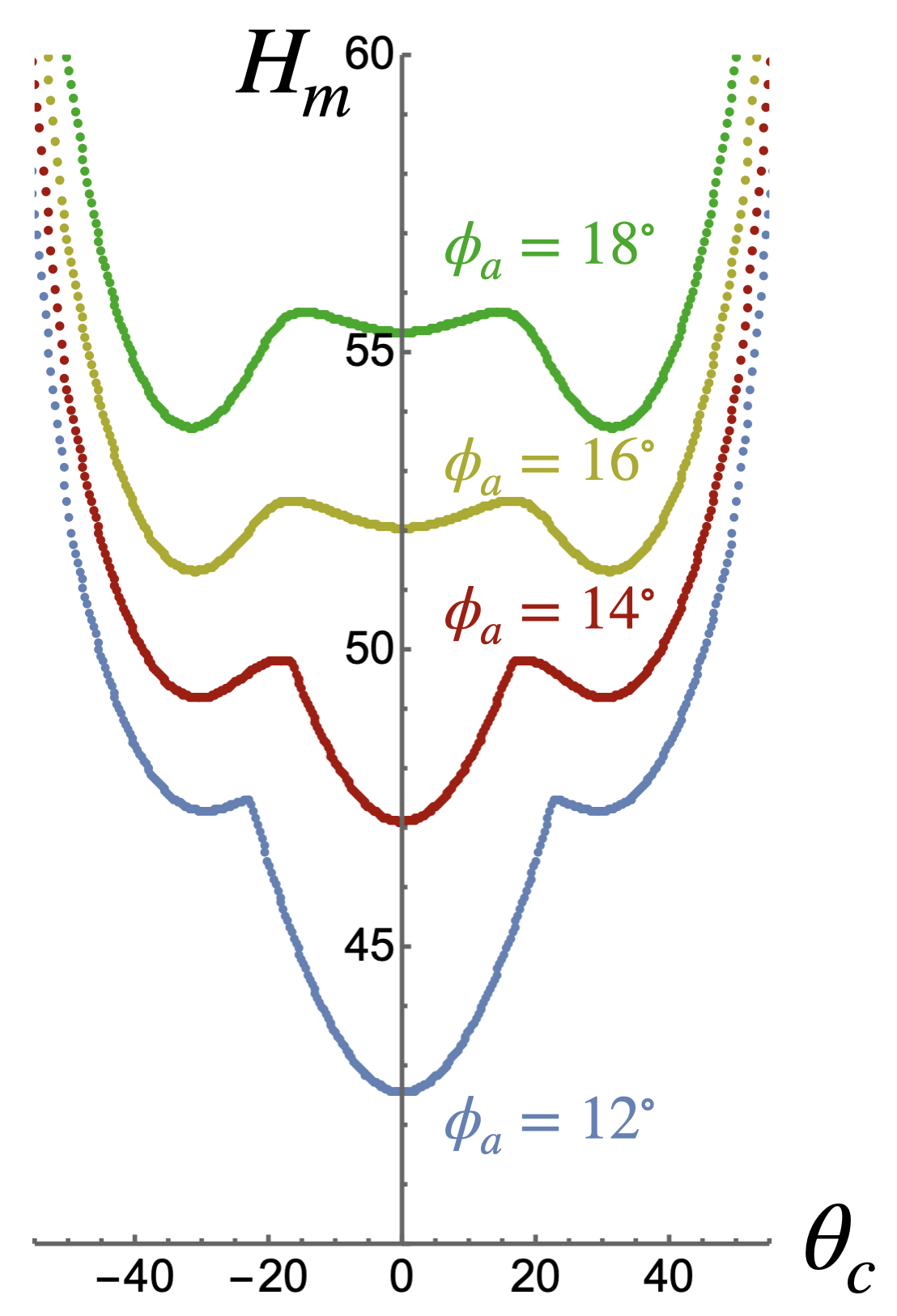}
    \caption{
    Theoretically computed \(H_m\) versus \(\theta_c\) for \(\phi_a=\{12^\circ,14^\circ,16^\circ,18^\circ\}\) obtained by minimizing the Landau free energy of Eq.~(\ref{F_metamagnetic}). We find good qualitative agreement with the experimental result presented in Fig.~\ref{fig:angles}(c). The nonzero parameters in Eq.~(\ref{F_metamagnetic}) used to make the plot are: \(\chi^{-1}_a=5.16\), \(\chi^{-1}_b=181\), \(\chi^{-1}_c=72.3\), \(\beta_{aa}=48.2\), \(\beta_{bb}=-688\), \(\beta_{cc}=181\), \(\beta_{ab}=-2168\), \(\delta_{bbb}=903\), \(\delta_{ccc}=1445\), \(\delta_{bbc}=108\), \(\delta_{bcc}=-1734\), \(\delta_{abb}=17616\),  \(\delta_{aab}=7046\),  and \(\delta_{abc}=22584\).
    %\textcolor{blue}{[MC: preliminary figure here for now]}  
    }
    \label{fig:theory}
\end{figure}

To describe the metamagnetic transition in UTe$_2$ theoretically, we use the Landau free energy~\cite{Yamada_metamagnetic,tony2024enhanced,qcl}
\begin{equation}
    \begin{gathered}
        \mathcal{F}=\frac{\chi^{-1}_{i}}{2}M_i^2+\frac{\beta_{ij}}{4} M_i^2M_j^2 +  \frac{\delta_{ijk}}{6}  M_i^2 M_j^2 M_k^2 -\mathbf{M\cdot H},
    \end{gathered}
\label{F_metamagnetic}
\end{equation}
where $i,j,k=a,b,c$ are spatial indices and parameters \(\chi^{-1}_{i}, \beta_{ij}\), and \(\delta_{ijk}\) determine the metamagnetic phase transition, which occurs when a non-trivial minimum of the free energy becomes a global minimum once \(H>H_m\).
We find a range of parameters for which the calculated metamagnetic transition lines in Fig.~\ref{fig:theory} qualitatively reproduce the measured transition lines shown in Fig.~\ref{fig:angles}(c). We note that the appearance of the ripples seems to be due to a proximity to an additional non-trivial local minimum in the free energy [Eq.~(\ref{F_metamagnetic})], which may lead to other observable effects that can be probed in future experiments, for example, those sensitive to fluctuations around this minimum~\cite{zambra2026giant}.

\section{Discussion}
\label{sec:discussion}
\subsection{Nature of the polarized paramagnetic state}
\label{sec:disc_pm}
Our finding of non-monotonic $H$ and $T$ dependencies of metamagnetism in UTe$_2$ underscores the complexity of this material's high-$H$ phase diagram. In our previous work~\cite{tony2024enhanced} we proposed that for $H < H_m$, the superconductive state that persists up to $\mu_0 H \approx$~34~T for \textbf{H} aligned close to $b$ is likely mediated by ferromagnetic-like fluctuations in the vicinity of $H_m$. Evidence favoring this scenario stems from the sensitivity to disorder of the angular extent of this superconducting phase~\cite{tony2024enhanced}, along with indirect measurements of the magnetic fluctuations from a magnetic entropy analysis~\cite{TokiwaPRB2024}, the findings of which concur with those from recent NMR experiments -- a directly sensitive probe of the local spin susceptibility~\cite{Kinjo23,Kinjo_SciAdv23,tokunaga2023longitudinal}. Notably, these NMR studies have observed a rotation of the local spin susceptibility from being predominantly along $a$ at low $H$ to aligning along $b$ as $H_m$ is approached, as one may expect for the scenario of triplet superconductivity mediated by ferromagnetic-like fluctuations in the vicinity of $H_m$ for large \textbf{H}~$\parallel b$.

However, the findings of the present study underline how little we understand about the microscopic properties of the polarized paramagnetic phase itself. A key challenge in studying UTe$_2$'s polarized paramagnetism lies in the necessity to access magnetic fields in excess of 34~T, precluding many scattering studies utilizing e.g. x-rays or neutrons that could map collective modes of magnetic excitations. Extraction magnetometry measurements in pulsed fields~\cite{Miyake2019,Miyake2021} have indicated that $H_m$ for \textbf{H}~$\parallel b$ is characterized by a sudden increase in the bulk magnetization of the material by an amount $\approx$ 0.5 $\mu_{\text{B}}$ per uranium (where $\mu_{\text{B}}$ is the Bohr magneton). A simple interpretation of this would be to assume that the polarized paramagnetic state is essentially ferromagnetic-like in character, with localized moments aligned along $b$. However, our observation of ripples in the metamagnetic transition surface at high $H$, accompanied by an anomalous temperature evolution of $H_m$ in the vicinity of these ripples, implies that the microscopic magnetic properties of UTe$_2$ at $H > H_m$ may be far more complex.

Beyond the zeroth-order scenario of ferromagnetic-like ordering, we speculate that the polarized paramagnet may instead exhibit some more complicated magnetic structure, for example ferrimagnetic or helimagnetic ordering. Such a scenario would not be inconsistent with the extraction magnetometry result of a sudden jump in the net magnetization (provided that, for some complex helimagnetic ordering, the moments are canted in such a manner to yield an overall net magnetization). Furthermore, the anomalous temperature dependence of $H_m$ we observed for \textbf{H} tilted to an orientation of $\phi_a = 18\degree$, $\theta_c = 0\degree$ (Figs.~\ref{fig:temps} and \ref{fig:3dtemps}) suggests that, rather than consisting of one single homogeneous phase as previously thought~\cite{lewin2023review}, perhaps the large region of the high-$H$ phase diagram generally referred to as `polarized paramagnetism' actually comprises one or more magnetic sub-phases. Such sub-phases could in principle exhibit distinctly different temperature evolutions. One possibility could be a form of helimagnetic ordering that is sensitive to the magnitude and orientation of \textbf{H}, yielding a slight rotation of the magnetic structure as a function of \textbf{H} and $T$, thereby giving the ripple-like structure and seemingly anomalous $T$ dependence of $H_m$ that we observed.

%Indeed, numerous studies have indicated that the uranium valence in UTe$_2$ lies between 3+ and 4+~\cite{Aoki_UTe2review2022,ute2valence_MPI24-PRR.6.033299}, while inelastic neutron scattering measurements at zero $H$ have reported fluctuating moments of magnitude 2.3(7) $\mu_{\text{B}}$ oriented along the $a$ direction~\cite{halloran2025npj}, which appears roughly consistent with both the bulk magnetization measured by SQUID magnetometry~\cite{Ran2019Science,ikeda2006old-ute2}. and the paramagnetic susceptibility of a free U 3+ (4+) ion, of 3.62 $\mu_{\text{B}}$ (3.58 $\mu_{\text{B}}$)~\cite{kindra2014magnetic}. Therefore, it is perhaps surprising why the jump in magnetization at $H_m$ is only 0.5~$\mu_{\text{B}}$. If the local uranium site moments were indeed larger for $H > H_m$ -- but partially compensated by antiparallel components from ferri or heli ordering -- this would provide a natural explanation as to why the jump at $H_m$ is relatively small.

We note that the anomalous temperature dependence of $H_m$ (Figs.~\ref{fig:temps} and \ref{fig:3dtemps}) within the ripple in the metamagnetic transition surface is located at high \textbf{H} tilted in the $b-a$ plane, i.e. with zero $H_c$ component ($\theta_c = 0\degree$). For $\mu_0 H \gtrapprox$~40~T, $b$ is the easy magnetic direction in UTe$_2$, while $c$ is the hard axis~\cite{Miyake2021}. Should there indeed exist some complex magnetic structure -- either within a homogeneous field-polarized state, or in tandem with one or more magnetic sub-phases spanning the phase landscape -- this observed sensitivity of $H_m(T)$ to zero and nonzero components of \textbf{H} along the hard $c$ direction perhaps provides a clue as to what the underlying high-$H$ magnetic structure may be.

These magnetic scenarios are not the only possibility. The metamagnetic transition in UTe$_2$ is likely accompanied by a reconstruction of the Fermi surface~\cite{NiuPhysRevResearch.2.033179}, and its itinerant aspects have been discussed in detail in \cite{weinberger2026metamagnetismute2rolesitinerancy}. From this perspective $H_m$ is also sensitive to the underlying electronic structure, including the anisotropic Zeeman response of the heavy quasiparticles, the Fermi surface topology and field-induced changes in band polarization. Each of these depends on the orientation of \textbf{H}, and could in principle produce non-monotonic structure in $H_m(\theta_c,\phi_a)$ within a single homogeneous polarized state, for instance if a Zeeman-split band edge is brought close to the Fermi level only for particular field orientations. The Landau free energy of Eq.~(\ref{F_metamagnetic}) does not distinguish between these pictures: it has the same form as that used to describe itinerant-electron metamagnetism~\cite{Yamada_metamagnetic}, in which case the additional local minimum responsible for the ripples would reflect structure in the quasiparticle density of states.

\subsection{\texorpdfstring{Thermodynamic implications of $H_m(T)$}{Thermodynamic implications of Hm(T)}}
\label{sec:disc_thermo}
The temperature dependence of $H_m$ carries direct thermodynamic information. Along the first-order portion of the metamagnetic phase boundary, the magnetic Clausius--Clapeyron relation gives $\mu_0\,\mathrm{d}H_m/\mathrm{d}T = -\Delta S/\Delta M$~\cite{Blundell2010thermal}, where $\Delta S = S_{\mathrm{PPM}} - S_{\mathrm{PM}}$ and $\Delta M = M_{\mathrm{PPM}} - M_{\mathrm{PM}}$ are the differences in entropy and in magnetization along \textbf{H} between the polarized paramagnetic (PPM) and paramagnetic (PM) states at $H_m$. Since the PPM state is the stable state above $H_m$, it must have the larger magnetization, so $\Delta M > 0$, as is observed experimentally~\cite{Miyake2019,Miyake2021}. The sign of $\mathrm{d}H_m/\mathrm{d}T$ therefore gives the sign of $\Delta S$ directly, without requiring knowledge of the size of $\Delta M$.

Outside the ripple, at $\phi_a = 18\degree$, $\theta_c = 20\degree$, $H_m$ decreases on warming (Fig.~\ref{fig:3dtemps}), implying $\Delta S > 0$, i.e.\ the polarized state has the higher entropy. Within the ripple, at $\phi_a = 18\degree$, $\theta_c = 0\degree$, $H_m$ instead increases on warming between 1.7~K and 6.0~K, implying $\Delta S < 0$: at this orientation the polarized state has a \textit{lower} entropy than the low-field paramagnetic state. The sign of the entropy change across the metamagnetic transition thus reverses between two orientations only $20\degree$ apart in $\theta_c$. The measurements in Fig.~\ref{fig:Tdep_new} show how this evolves across the ripple. At $\phi_a = 16\degree$, $\theta_c = 0\degree$ the increase of $H_m$ on warming, and hence the entropy deficit of the polarized state, is weaker, while at $\phi_a = 19\degree$, $\theta_c = 6\degree$ it is marginal and confined to the lowest temperatures. The increase of $H_m$ below 6~K at $\phi_a = 18\degree$, $\theta_c = 0\degree$ is observed in both rising and falling field sweeps (Fig.~\ref{fig:updw_Tdep}(a)). A temperature-dependent hysteresis would shift these in opposite directions, so the sign of $\Delta S$ is not an artifact of hysteresis. Provided the transition remains first order and $\Delta M$ remains finite, the maximum of $H_m(T)$ at $\approx 6$~K marks the temperature at which the two states have equal entropy, above which $\Delta S > 0$, as at $\theta_c = 20\degree$, before the transition broadens into a crossover. The magnitude of $\Delta S$ cannot be obtained from our data alone: this would require measurements of $\Delta M(T)$ and of the equilibrium phase boundary at this orientation.

A lower entropy of the polarized state within the ripple could indicate that, at this orientation, the high-field state is the more ordered of the two, as would be expected if additional magnetic order develops above $H_m$. Alternatively, the low-field paramagnetic state may carry an enhanced entropy from magnetic fluctuations at this field orientation. Our measurements cannot distinguish between these possibilities. Magnetocaloric measurements in pulsed fields at this orientation would provide a direct test, since the latent heat on crossing $H_m$, separated from hysteretic heating by comparing rising and falling sweeps, gives both the sign and the size of $\Delta S$.

\subsection{Possible connection to SC3 superconductivity}
\label{sec:disc_sc3}
We speculate that the possible magnetic substructure discussed in Sec.~\ref{sec:disc_pm} may play an important role in the formation of the high-field SC3 superconducting state. At temperatures immediately above the $T_c$ of SC3, Planckian linear-in-$T$ resistivity has been observed~\cite{weinberger2025strangemetallicityencompasseshigh}, pointing to a quantum critical origin of the superconductive pairing interaction. However, somewhat puzzlingly, although it has been established that the metamagnetic transition surface terminates at a quantum critical phase boundary~\cite{qcl}, this is located relatively far from the magnetic field orientation at which SC3 is maximal in its $T$ and $H$ extent. This suggests that the suppression of some other -- previously unidentified -- phase to a quantum critical point likely drives SC3. Our observation of ripples in the metamagnetic transition surface at high $\phi_a$, relatively far from the region in which SC3 is strongest, provides a likely candidate for this unidentified additional state. Future measurements tracking how these features evolve under rotation towards the domain of SC3 will provide valuable empirical impetus towards resolving the microscopic origin of this exotic superconducting state.

\section{Summary}
\label{sec:summary}
In summary, we mapped the metamagnetic transition surface of UTe$_2$ for rotations of \textbf{H} through the polar angle $\theta_c$ at fixed azimuthal inclinations of $\phi_a$. We observed a ripple-like non-monotonic dependence of the metamagnetic transition field $H_m$ on $\theta_c$ at high $H$ and high $\phi_a$. Within this ripple the melting of the metamagnetic transition follows an anomalous non-monotonic temperature profile upon warming. These findings indicate that the magnetic properties of the high-$H$ polarized paramagnetic state in UTe$_2$ are considerably more complex than previously considered.

%\clearpage

\begin{acknowledgments}
We gratefully acknowledge stimulating discussions with Theo Weinberger and Alex Hickey. This project was supported by the EPSRC of the UK (grant nos. EP/X011992/1 \& EP/Z533695/1). Crystal growth and characterization were performed in MGML (mgml.eu), which is supported within the program of Czech Research Infrastructures (project no. LM2023065). We acknowledge financial support by the Czech Science Foundation (GACR), project No. 22-22322S. This work was performed in part at the Aspen Center for Physics, which is supported by National Science Foundation grant PHY-2210452. The work at UW-Madison (D.S.) was financially supported by the National Science Foundation, Quantum Leap Challenge Institute for Hybrid Quantum Architectures and Networks Grant No. OMA-2016136. D.V.C. acknowledges financial support from the National High Magnetic Field Laboratory through a Dirac Fellowship, which is funded by the National Science Foundation (Grant No. DMR-2128556) and the State of Florida. A.G.E. acknowledges support from Sidney Sussex College (University of Cambridge).
\end{acknowledgments}

% ======================================================================
%  APPENDIX: rising vs falling field sweeps only (the dashed-curve figures).
%  A single appendix, so \appendix* (APS style: "APPENDIX", no letter).
%  The other former SI figures now sit in the main text (Sec. III).
%  The two panels of old SI Fig. S6 are stacked in one single-column figure
%  (source in Appendix_Figures/split_source/stack_Tdep_updw.tex).
% ======================================================================
\appendix*

% ----------------------------------------------------------------------
\section{Rising and falling magnetic field sweeps}
\label{app:sweeps}
\begin{figure}[h!]
    \centering
    \includegraphics[width=0.8\linewidth]{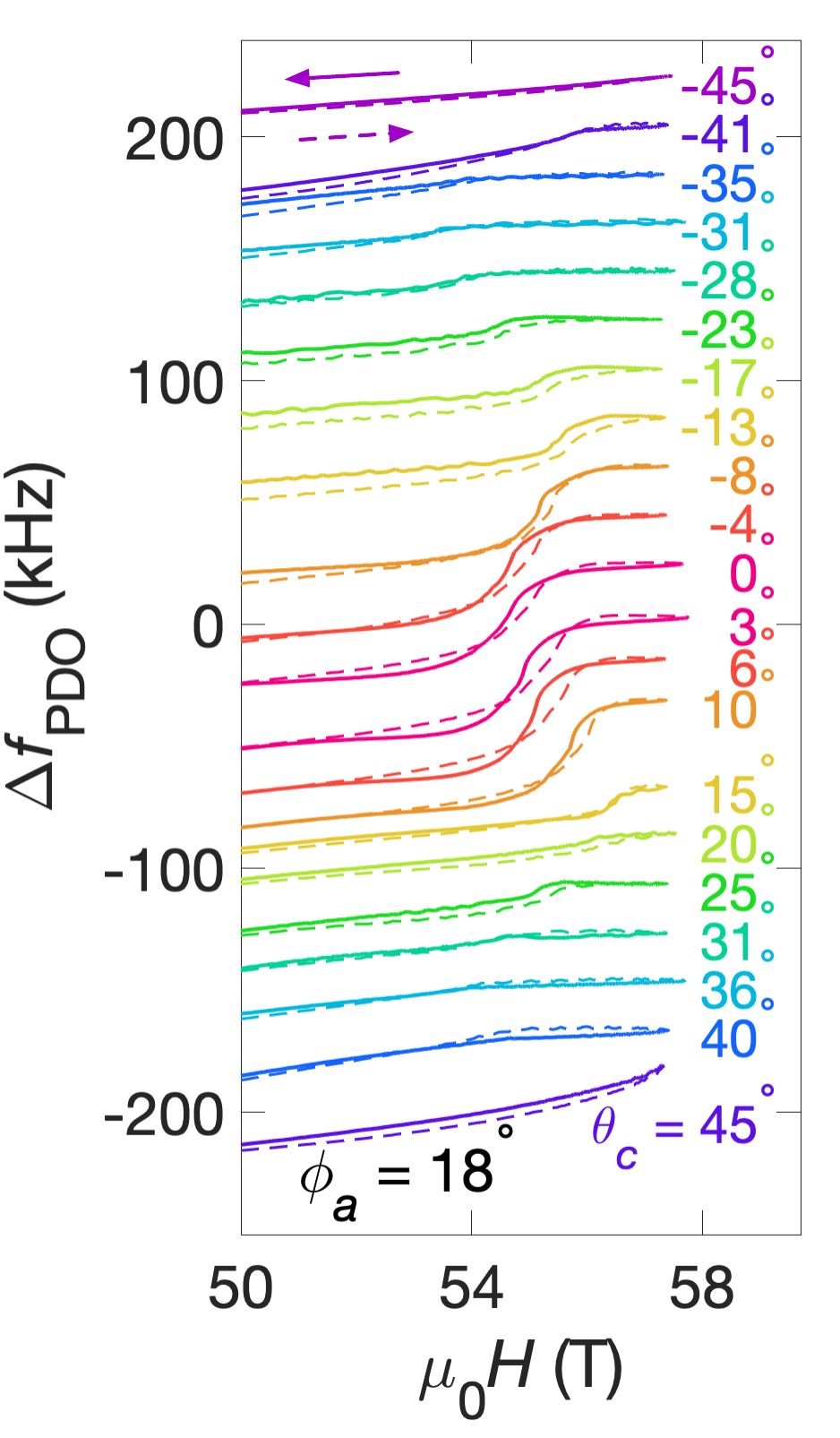}
    \caption{Here we plot the measured PDO frequency change for the rising (dashed lines) and falling (solid lines) portions of the magnetic field pulse. In Fig.~\ref{fig:angles} we plot just the falling section of the pulse, as this lasts for a longer duration than the rising portion, thereby giving better resolution. The ripple profile is observed for both rising and falling field, with signatures of some hysteresis at low $\theta_c$.}
    \label{fig:updw_crack}
\end{figure}

\begin{figure}[h!]
    \centering
    \includegraphics[width=0.85\linewidth]{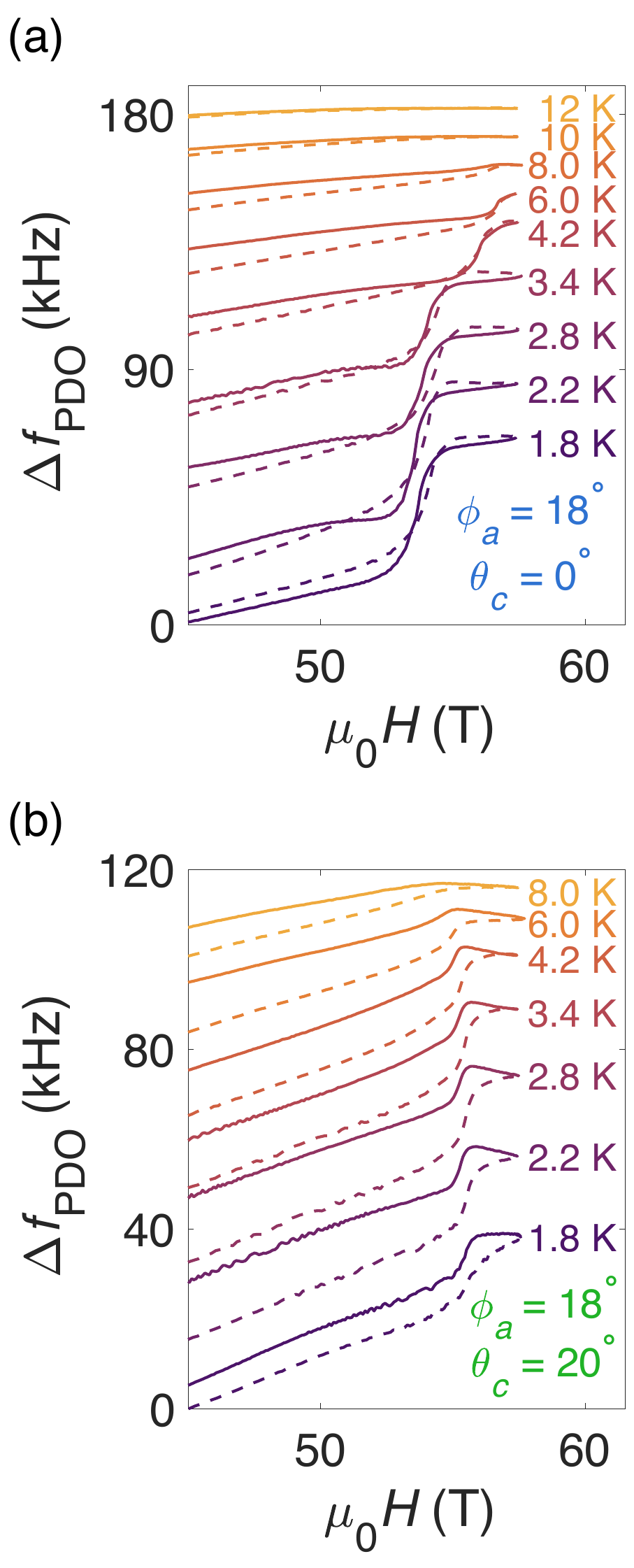}
    \caption{Temperature dependence of PDO signal for (a) $\phi_a = 18\degree$, $\theta_c = 0\degree$ and (b) $\phi_a = 18\degree$, $\theta_c = 20\degree$. Data are reproduced from Figs.~\ref{fig:temps}(a) and \ref{fig:temps}(c), and here we also plot the rising portion of the magnetic field pulse (dashed curves). The overall profile of $\Delta f_{\text{PDO}}$ tracks very similarly for both the rising and falling sections of the magnetic field pulse.}
    \label{fig:updw_Tdep}
\end{figure}

\begin{figure}[h!]
    \centering
    \includegraphics[width=0.70\linewidth]{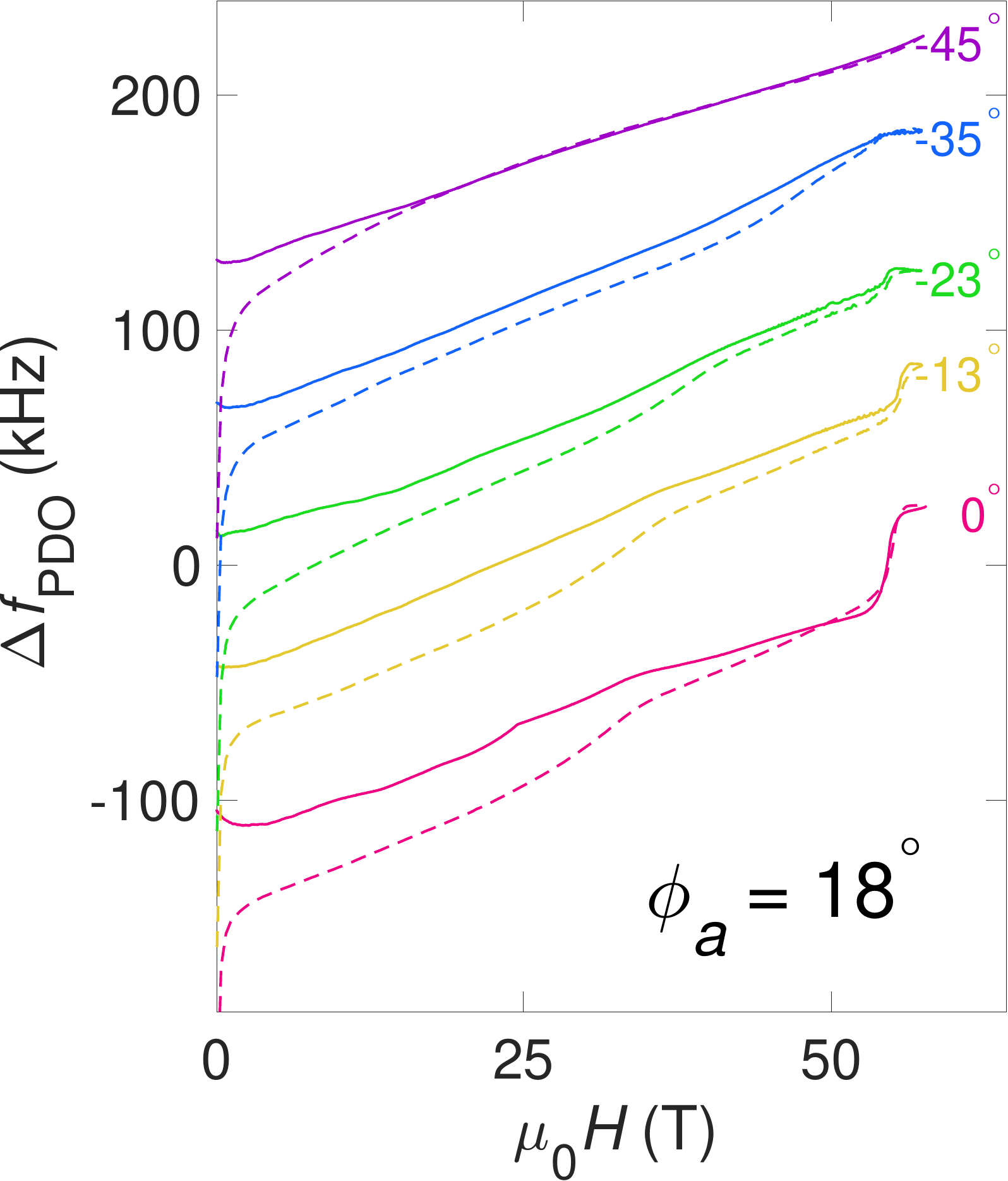}
    \caption{Representative PDO data over the full field range, for rising (dashed) and falling (solid) portions of the field sweep. $\phi_a = 18\degree$ for all data, with $\theta_c$ values as indicated.}
    \label{fig:fullrange}
\end{figure}

% ----------------------------------------------------------------------
% In Figure~\ref{fig:updw_crack} we plot the measured PDO frequency change for the rising (dashed lines) and falling (solid lines) portions of the magnetic field pulse. In the main text we plot just the falling section of the pulse, as this lasts for a longer duration than the rising portion thereby giving better resolution.
%We can see that up sweeps (dashed line) matches well with the down sweeps (solid line) and also tracks the ripple feature in field - angle phase diagram and non-monotonic temperature dependence of metamangetic (MM) transition location $H_M$. 
Figure~\ref{fig:updw_crack} compares the rising and falling portions of the magnetic field pulse for the $\phi_a = 18\degree$ rotation shown in Fig.~\ref{fig:angles}, and Fig.~\ref{fig:updw_Tdep} compares them for the temperature dependence shown in Fig.~\ref{fig:temps}. Representative data over the full field range are shown in Fig.~\ref{fig:fullrange}.

% \begin{figure}[h!]
%     \centering
%     \includegraphics[width=\linewidth]{Appendix_Figures/SuppFig_3DThetaPhiPD_SC3_smooth.pdf}
%     \caption{Same rendering as Fig.~\ref{fig:3D}(b), plotted here in enlarged format to more clearly show how the toroidal domain of SC3 relates to the metamagnetic ripples. Purple superconducting region taken from~\cite{qcl}, which matches closely with that reported in \cite{lewin2025halo}.}
%     \label{fig:3DPD}
% \end{figure}

\clearpage
\bibliography{UTe2}
\end{document}